\documentclass[epj]{webofc}
\usepackage[varg]{txfonts}   

\usepackage{amsmath,amsfonts,amssymb,bm}
\usepackage{dsfont}
\usepackage{graphicx}
\usepackage{color}
\usepackage{soul}

\usepackage[mathscr,scaled=1.15]{urwchancal}
\DeclareFontFamily{OT1}{pzc}{}
\DeclareFontShape{OT1}{pzc}{m}{it}%
{<-> s * [1.15] pzcmi7t}{}
\DeclareMathAlphabet{\mathpzc}{OT1}{pzc}{m}{it}

\definecolor{purple}{rgb}{0.5,0,0.5}
\definecolor{blue}{rgb}{0.0,0,0.9}
\definecolor{prdblue}{rgb}{0.133,0.118,0.498}

\wocname{EPJ Web of Conferences}
\woctitle{CONF12}
\begin{document}
\selectlanguage{english}
\title{Emergent phenomena and partonic structure in hadrons}
%
%

\author{Craig D. Roberts\inst{1}\fnsep\thanks{\email{cdroberts@anl.gov}}
and C\'edric Mezrag\inst{1}\fnsep\thanks{\email{cmezrag@anl.gov}}
}

\institute{Physics Division, Argonne National Laboratory, Argonne IL 60439, USA
}

\abstract{%
Modern facilities are poised to tackle fundamental questions within the Standard Model, aiming to reveal the nature of confinement, its relationship to dynamical chiral symmetry breaking (DCSB) -- the origin of visible mass -- and the connection between these two, key emergent phenomena.  There is strong evidence to suggest that they are intimately connected with the appearance of momentum-dependent masses for gluons and quarks in QCD, which are large in the infrared: $m_g \sim 500\,$MeV and $M_q\sim 350\,$MeV.  DCSB, expressed in the dynamical generation of a dressed-quark mass, has an enormous variety of verifiable consequences, including an enigmatic result that the properties of the (almost) massless pion are the cleanest expression of the mechanism which is responsible for almost all the visible mass in the Universe.  This contribution explains that these emergent phenomena are expressed with particular force in the partonic structure of hadrons, \emph{e.g}.\ in valence-quark parton distribution amplitudes and functions, and, consequently, in numerous hadronic observables, so that we are now in a position to exhibit the consequences of confinement and DCSB in a wide range of hadron observables, opening the way to empirical verification of their expression in the Standard Model.
}
\maketitle
\section{Whence mass?}
\label{intro}
%
Classical chromodynamics (CCD) is a non-Abelian local gauge field theory.  As with all such theories formulated in four spacetime dimensions, no length-scale exists in the absence of Lagrangian masses for the fermions.  There is no dynamics in a scale-invariant theory, only kinematics: the theory looks the same at all length-scales and hence there can be no clumps of anything.  Bound-states are therefore impossible and, accordingly, our Universe cannot exist.  Spontaneous symmetry breaking, as realised via the Higgs mechanism, does not solve this problem because normal matter is constituted from light-quarks, $u$ and $d$, and the masses of the neutron and proton, the kernels of all visible matter, are roughly 100-times larger than anything the Higgs can produce in connection with $u$- and $d$-quarks.  Consequently, the question of how did the Universe come into being is inseparable from the questions of how does a mass-scale appear and why does it have the value we observe?

Energy and momentum conservation in a quantum field theory is a consequence of spacetime translational invariance, one of the family of Poincar\'e transformations.  Consequently,
\begin{equation}
\label{ConsEP}
\partial_\mu T_{\mu\nu} = 0 \,,
\end{equation}
where $T_{\mu\nu}$ is the energy-momentum tensor in CCD, which can always be made symmetric \cite{BELINFANTE1940449}.   A global scale translation is implemented via:
\begin{equation}
\label{scaleT}
x  \to x^\prime = {\rm e}^{-\sigma}x\,, \quad
A_\mu^a(x)  \to A_\mu^{a\prime}(x^\prime) = {\rm e}^{-\sigma} A_\mu^a({\rm e}^{-\sigma}x ) \,, \quad
q(x)  \to q^\prime(x^\prime) = {\rm e}^{- (3/2) \sigma} q({\rm e}^{-\sigma}x )\,,
\end{equation}
where $A_\mu^a(x)$, $q(x)$ are the gluon and quark fields.  The Noether current associated with this transformation is the dilation current:
${\cal D}_\mu = T_{\mu\nu} x_\nu$.
In the absence of fermion masses, the classical action is invariant under Eqs.\,\eqref{scaleT}, \emph{i.e}.\ the theory is scale invariant, and hence
\begin{equation}
\partial_\mu {\cal D}_\mu  = 0  = [\partial_\mu T_{\mu\nu} ] x_\nu + T_{\mu\nu} \delta_{\mu\nu}  = T_{\mu\mu}\,,
\label{SIcQCD}
\end{equation}
where the last equality follows from Eq.\,\eqref{ConsEP}: the energy-momentum tensor is traceless in a scale invariant theory.

Massless CCD is meaningless for many reasons; amongst them the fact that strong interactions in the Standard Model are empirically known to be characterised by a large mass-scale, $\Lambda_{\rm QCD}\approx 0.2\,$GeV \cite{Agashe:2014kda}.  In quantising the theory, regularisation and renormalisation of (ultraviolet) divergences introduces a mass-scale.  This is ``dimensional transmutation'': mass-dimensionless quantities become dependent on a mass-scale, and this entails the violation of Eq.\,\eqref{SIcQCD}, \emph{i.e}. the appearance of the chiral-limit ``trace anomaly'':
\begin{equation}
\label{SIQCD}
T_{\mu\mu} = \beta(\alpha(\zeta))  \tfrac{1}{4} G^{a}_{\mu\nu}G^{a}_{\mu\nu} =: \Theta_0 \,,
\end{equation}
where $\beta(\alpha)$ is the $\beta$-function of quantum chromodynamics (QCD), $\zeta$ is the renormalisation scale, $G^{a}_{\mu\nu}$ is the gluon field-strength tensor, and this expression assumes the chiral limit for all current-quarks.

It is worth highlighting that the appearance of a trace anomaly has nothing to do with the non-Abelian nature of QCD.  Indeed, quantum electrodynamics (QED) also possesses a trace anomaly.  However, QED is nonperturbatively undefined: four-fermion operators become relevant in strong-coupling QED and must be included in order to obtain a well-defined (albeit trivial) continuum limit (see, \emph{e.g}.\, Refs.\,\cite{Rakow:1990jv, Reenders:1999bg, Akram:2012jqS}).  As a consequence, QED does not have a chiral limit.  The QED trace anomaly is only meaningful in perturbation theory and its scale is determined by the Higgs mechanism.

In the presence of nonzero current-quark masses, Eq.\,\eqref{SIQCD} becomes
\begin{equation}
\Theta:= T_{\mu\mu} = \tfrac{1}{4} \beta(\alpha(\zeta)) G^{a}_{\mu\nu}G^{a}_{\mu\nu}
+ [1+\gamma(\alpha(\zeta))]\sum_f m_f^\zeta \, \bar q_f q_f\,,
\end{equation}
where $m_f^\zeta$ are the current-quark masses and $[1+\gamma(\alpha)]$ is the analogue for the dressed-quark running-mass of $\beta(\alpha)$ for the running coupling.  (In fact, $\gamma(\alpha)$ is the anomalous dimension of the current-quark mass in QCD.)
It is notable that in the massive-case the trace anomaly is not homogeneous in the running coupling, $\alpha(\zeta)$.  Consequently, renormalisation-group-invariance does not entail form invariance of the right-hand-side (rhs) \cite{Tarrach:1981bi}.  This is important because discussions typically assume (perhaps implicitly) that all operators and identities are expressed in a partonic basis, \emph{viz}.\ using simple field operators that can be renormalised perturbatively, in which case the hadronic state-vector represents an extremely complicated wave function.  That perspective is not valid at renormalisation scales $\zeta \lesssim m_p$, where $m_p$ is the proton mass; and this is where a metamorphosis from parton-basis to quasiparticle-basis may occur: under reductions in resolving scale, $\zeta$, light partons evolve into heavy dressed-partons, corresponding to complex superpositions of partonic operators; and using these dressed-parton operators, the wave functions can be expressed in a relatively simple form \cite{Roberts:2016vyn}.
\section{Magnitude of the scale anomaly: mass and masslessness}
%
%
Simply knowing that a trace anomaly exists does not deliver a great deal: it only indicates that there is a mass-scale.  The crucial issue is whether or not one can compute and/or understand the magnitude of that scale.
%
One can certainly measure the size of the scale anomaly, for consider the expectation value of the energy-momentum tensor in the proton (\emph{e.g}., Ref.\,\cite{Kharzeev:1995ij}):
\begin{equation}
\label{EPTproton}
\langle p(P) | T_{\mu\nu} | p(P) \rangle = - P_\mu P_\nu\,,
\end{equation}
where the rhs follows from the equations-of-motion for an asymptotic one-particle proton state.  At this point it is clear that, in the chiral limit,
\begin{equation}
\label{anomalyproton}
\langle p(P) | T_{\mu\mu} | p(P) \rangle  = - P^2  = m_p^2 = \langle p(P) |  \Theta_0 | p(P) \rangle\,;
\end{equation}
namely, there is a clear sense in which it is possible to say that the entirety of the proton mass is produced by gluons.  The trace anomaly is measurably large; and that property must logically owe to gluon self-interactions, which are also responsible for asymptotic freedom.  This is a valid conclusion.  After all, what else could be responsible for a mass-scale?  QCD is all about gluon self-interactions; and it's gluon self-interactions that (potentially) enable one to rigorously (nonperturbatively) define the expectation value in Eq.\,\eqref{anomalyproton}.  On the other hand, it's only a sensible conclusion when the operator and the wave function are defined at a resolving-scale $\zeta \gg m_p$.

There is also another issue, which can be exposed by returning to Eq.\,\eqref{EPTproton} and replacing the proton by the pion:
\begin{equation}
\label{EPTpion}
\langle \pi(q) | T_{\mu\nu} | \pi(q) \rangle = - q_\mu q_\nu\,.
\end{equation}
Then, in the chiral limit:
\begin{align}
\label{anomalypion}
& \langle \pi(q) |  \Theta_0 | \pi (q) \rangle = 0
\end{align}
because the pion is a massless Nambu-Goldstone mode.  Equation\,\eqref{EPTpion} could mean that the scale anomaly vanishes trivially in the pion state, \emph{viz}.\ that gluons and their self-interactions have no impact within a pion because each term in the expression of the operator vanishes when evaluated in the pion. However, that is a difficult way to achieve Eq.\,\eqref{anomalypion}.  It is easier to imagine that Eq.\,\eqref{anomalypion} owes to cancellations between different operator-component contributions.  Of course, such precise cancellation should not be an accident.  It could only arise naturally because of some symmetry and/or symmetry-breaking pattern; and, as will be argued, that is the manner by which Eq.\,\eqref{anomalypion} is realised.

Equations\,\eqref{anomalyproton} and \eqref{anomalypion} present a quandary, which highlights that no understanding of the origin of the proton's mass can be complete unless it simultaneously explains the meaning of Eq.\,\eqref{anomalypion}.  Given that a massless particle doesn't have a rest-frame, any approach based on a rest-frame decomposition of the energy-momentum tensor (\emph{e.g}.\ Ref.\,\cite{Ji:1995sv}) cannot readily be useful in this dual connection.

It is worth recalling here that CCD is still a non-Abelian local gauge theory.  Consequently, the concept of local gauge invariance persists.  However, without a mass-scale there is no confinement.  For example, three quarks can be prepared in a colour-singlet combination and colour rotations will keep the three-body system colour neutral; but the quarks involved need not have any proximity to one another.  Indeed, proximity is meaningless because all lengths are equivalent in a scale invariant theory.  Hence, the question of ``Whence mass''? is equivalent to ``Whence a mass-scale?'', which is equivalent to ``Whence a confinement scale?''.  Thus, understanding the origin, Eq.\,\eqref{anomalyproton}, and absence, Eq.\,\eqref{anomalypion}, of mass in QCD is quite likely inseparable from the task of understanding confinement; and existence, alone, of a scale anomaly answers neither question.

A point has now been reached from which a set of basic questions can meaningfully be posed and, perhaps, answered.  Namely, what is origin of mass in our Universe; what is the nature of confinement in real-QCD, \emph{i.e}.\ in the presence of active light-quarks; and how are they connected?  Of great importance, too, is the associated consideration: how can any answers, conjectures and/or conclusions be empirically verified?  Physics is an empirical science and no explanation is physical unless it can realistically be subjected to experimental validation.

\section{Confinement and DCSB}
QCD is characterised by two emergent phenomena: confinement and dynamical chiral symmetry breaking (DCSB).  They have far-reaching consequences, expressed with great force in the character of the pion, and pion properties, in turn, suggest that confinement and DCSB are intimately connected.  Indeed, since the pion is both a Nambu-Goldstone boson and a quark-antiquark bound-state, it holds a unique position in Nature and, consequently, developing an understanding of its properties is critical to revealing some very basic features of the Standard Model.\\[-2ex]

\hspace*{2em}\parbox[t]{0.85\textwidth}{\textit{Confinement Hypothesis}: Colour-charged particles cannot be isolated and therefore cannot be directly observed.  They clump together in colour-neutral bound-states.}\\

Confinement seems to be an empirical fact; but a mathematical proof is lacking.  Partly as a consequence, the Clay Mathematics Institute offered a ``Millennium Problem'' prize of \$1-million for a proof that $SU_c(3)$ gauge theory is mathematically well-defined \cite{Jaffe:Clay}, one necessary consequence of which will be an answer to the question of whether or not the confinement conjecture is correct in pure-gauge QCD.  There is a problem with this, however: no reader of this article can be described within pure-gauge QCD.  The presence of quarks is essential to understanding all known visible matter, so a proof of confinement that deals only with pure-gauge QCD is chiefly irrelevant to our Universe.  We exist because Nature has supplied two light quarks and those quarks combine to form the pion, which is unnaturally light ($m_\pi < \Lambda_{\rm QCD})$ and hence very easily produced.  This point can be sharpened by noting that one aspect of the Yang-Mills millennium problem is to prove that pure-gauge QCD possesses a mass-gap $\Delta>0$.  There is strong evidence supporting this conjecture, found especially in the fact that numerical simulations of lattice-regularised QCD (lQCD) predict $\Delta \gtrsim 1.5\,$GeV \cite{McNeile:2008sr}.  However, with $\Delta^2/m_\pi^2 \gtrsim 100$, can the mass-gap in pure Yang-Mills theory really play any role in understanding confinement when DCSB, likely driven by the same dynamics, ensures the existence of an almost-massless strongly-interacting excitation in our Universe?  If the answer is not \emph{no}, then it must at least be that one cannot claim to provide a pertinent understanding of confinement without simultaneously explaining its connection with DCSB.  The pion must play a critical role in any explanation of confinement in the Standard Model; and therefore, as others have already noted \cite{Casher:1979vw, Banks:1979yr}, any discussion of confinement that omits reference to the pion's role is \emph{practically} \emph{irrelevant}.

This perspective is canvassed elsewhere \cite{Cloet:2013jya, Roberts:2015lja, Horn:2016rip}.  It can be used to argue that: the so-called ``vacuum condensates'' are actually wholly contained within hadrons \cite{Brodsky:2009zd, Brodsky:2010xf, Chang:2011mu, Brodsky:2012ku}; and the potential between infinitely-heavy quarks measured in numerical simulations of quenched lQCD -- the so-called static potential \cite{Wilson:1974sk} -- is disconnected from the question of confinement in our Universe.  The latter follows because light-particle creation and annihilation effects are essentially nonperturbative in QCD, so it is impossible in principle to compute a quantum mechanical potential between two light quarks \cite{Bali:2005fu, Prkacin:2005dc, Chang:2009ae}.
It follows that the flux tube measured in numerical simulations of lQCD with static quarks has no relevance to confinement in the light-quark realm of QCD.
Moreover, there is zero knowledge of the strength or extension of a flux tube between a static-quark and any light-quark because it is impossible even to define such a flux tube; and it is doubly impossible to define a flux-tube between a light-quark source and light-quark sink.
Hence, since the vast bulk of visible matter is constituted from light valence quarks, with no involvement of even an accessible heavy quark, the flux tube picture is not the correct paradigm for confinement in hadron physics.

An alternative realisation associates confinement with dramatic, dynamically-driven changes in the analytic structure of QCD's propagators and vertices.  That leads coloured $n$-point functions to violate the axiom of reflection positivity and hence forces elimination of the associated excitations from the Hilbert space associated with asymptotic states \cite{GJ81}.  This is certainly a sufficient condition for confinement \cite{Stingl:1985hx, Krein:1990sf, Hawes:1993ef, Roberts:1994dr}. Moreover, via this mechanism, it is achieved as the result of an essentially dynamical process.  It is known that quarks acquire a running mass distribution in QCD \cite{Bhagwat:2003vw, Bowman:2005vx, Bhagwat:2006tu}.  This is also true of gluons (see, \emph{e}.\emph{g}.\ Refs.\,\cite{Aguilar:2008xm, Aguilar:2009nf, Boucaud:2011ugS, Pennington:2011xs, Ayala:2012pb, Binosi:2014aea, Binosi:2016xxu}); and both are large at infrared (IR) momenta.  The generation of these masses leads to the emergence of a length-scale $\varsigma \approx 0.5\,$fm, whose existence and magnitude is evident in all existing studies of dressed-gluon and -quark propagators, and which characterises the dramatic change in their analytic structure that we have just described.  In models based on such features \cite{Stingl:1994nk}, once a gluon or quark is produced, it begins to propagate in spacetime; but after each ``step'' of length $\varsigma$, on average, an interaction occurs so that the parton loses its identity, sharing it with others.  Finally a cloud of partons is produced, which coalesces into colour-singlet final states.  This picture of parton propagation, hadronisation and confinement can be tested in experiments at modern and planned facilities \cite{Accardi:2009qv, Dudek:2012vr, Accardi:2012qutP}.

The connection between the existence of Nambu-Goldstone modes and a dynamically-generated dressed-quark mass is detailed in Ref.\,\cite{Horn:2016rip}.  The propagation of gluons, too, is described by a gap equation; and its solution shows that gluons are cannibals: they are a particle species whose members become massive by eating each other!  The associated gluon mass function, $m_g(k^2)$, is monotonically decreasing with increasing $k^2$ and recent work \cite{Binosi:2014aea} has established that
\begin{equation}
\label{gluonmassEq}
m_g(k^2=0) \approx 0.5\,{\rm GeV}.
\end{equation}
The value of the mass-scale in Eq.\,\eqref{gluonmassEq} is \emph{natural} in the sense that it is commensurate with but larger than the value of the dressed light-quark mass function at far infrared momenta: $M(0)\approx 0.3\,$GeV.  Moreover, the mass term appears in the transverse part of the gluon propagator, hence gauge-invariance is not tampered with; and the mass function falls as $1/k^2$ for $k^2\gg m_g(0)$ (up to logarithmic corrections), so the gluon mass is invisible in perturbative applications of QCD: it has dropped to less-than 5\% of it's infrared value by $k^2=4\,$GeV$^2$.

Gauge boson cannibalism presents a new physics frontier within the Standard Model.  Asymptotic freedom means that the ultraviolet behaviour of QCD is controllable.  At the other extreme, dynamically generated masses for gluons and quarks entail that QCD creates its own infrared cutoffs.  Together, these effects eliminate both the infrared and ultraviolet problems that typically plague quantum field theories and thereby make reasonable the hope that QCD is nonperturbatively well defined, \emph{viz}.\ that the millennium problem \cite{Jaffe:Clay} does have a solution.  The presence of dynamically-generated gluon and quark mass-scales must have many observable consequences, too, and hence can be checked experimentally.  For example, one may plausibly conjecture that dynamical generation of an infrared gluon mass-scale leads to saturation of the gluon parton distribution function at small Bjorken-$x$ within hadrons.  This could be checked via computations of gluon distribution functions, using such solutions of the gluon gap equation in hadron bound-state equations.  The possible emergence of this phenomenon stirs great interest; and it is a key motivation in plans to construct an electron ion collider (EIC) that would be capable of producing a precise empirical understanding of collective behaviour amongst gluons \cite{Accardi:2012qutP}.

A well-designed EIC could also provide a wealth of information and hence knowledge about the valence region, \emph{e.g}.\ exposing a marked difference in the gluon content of $\pi$- and $K$ mesons \cite{Chen:2016sno} by measuring their valence-quark distribution functions.  This would add greatly to our understanding of the \emph{structure} of Nambu-Goldstone bosons, which are far more than the simple pointlike modes described in extant textbooks.  (See, \emph{e.g}.\ Sec.\,\ref{secEnigma}.)

In order to practically address the questions posed above, one must possess detailed information about the interactions between quarks and gluons at all momentum scales so that the continuum bound-state problem can be solved.  There are two common methods for determining this information: the top-down approach, which works toward an \textit{ab initio} computation of the interaction via direct analysis of the gauge-sector gap equations; and the bottom-up scheme, which aims to infer the interaction by fitting data within a well-defined truncation of those equations in the matter sector that are relevant to bound-state properties.  These two approaches have recently been united \cite{Binosi:2014aea} by a demonstration that the renormalisation-group-invariant running-interaction predicted by contemporary analyses of QCD's gauge sector coincides with that required in order to describe ground-state hadron observables. This is a remarkable stride, given that there had previously been no serious attempt at communication between practitioners from the top-down and bottom-up hemispheres of continuum-QCD.  It bridges a gap that had lain between nonperturbative continuum-QCD and the \emph{ab initio} prediction of bound-state properties.  Critically, this advance was only made possible following an appreciation of just how important the dressed--gauge-boson--quark vertex is to DCSB, and vice versa, which had grown over many years \cite{Binosi:2016wcx}.

\section{Enigma of mass}
\label{secEnigma}
As noted above, the remarkably low mass of the pion, just one-fifth of that which quantum mechanics would lead one to expect, has its origin in DCSB.  In quantum field theory the pion's structure is described by a Bethe-Salpeter amplitude (here $k$ is the relative momentum between the valence-quark and -antiquark constituents, and $P$ is their total momentum):
\begin{equation}
\Gamma_{\pi}(k;P) = \gamma_5 \left[
i E_{\pi}(k;P) + \gamma\cdot P F_{\pi}(k;P)  + \gamma\cdot k \, G_{\pi}(k;P) - \sigma_{\mu\nu} k_\mu P_\nu H_{\pi}(k;P)
\right],
\label{genGpi}
\end{equation}
which is simply related to an object that would be the pion's Schr\"odinger wave function if a nonrelativistic limit were appropriate.  In QCD, if, and only if, chiral symmetry is dynamically broken, then in the chiral limit \cite{Maris:1997hd, Qin:2014vya, Binosi:2016rxz}:
\begin{equation}
\label{gtrE}
f_\pi E_\pi(k;0) = B(k^2)\,,
\end{equation}
where $f_\pi$ is the pion's leptonic decay constant, a directly measurable quantity that connects the strong and weak interactions, and the rhs is a scalar function in the dressed-quark propagator.  This identity is miraculous.  It means that the two-body problem is solved, almost completely, without lifting a finger, once the solution to the one body problem is known.  Eq.\,\eqref{gtrE} is a quark-level Goldberger-Treiman relation.  It is also the most basic expression of Goldstone's theorem in QCD, \emph{viz}.\\[-3ex]

\centerline{\parbox{0.85\textwidth}{\flushleft \emph{Goldstone's theorem is fundamentally an expression of equivalence between the one-body problem and the two-body problem in QCD's colour-singlet pseudoscalar channel.}}}

\medskip

\hspace*{-\parindent}Consequently, pion properties are an almost direct measure of the dressed-quark mass function; and the reason a pion is massless in the chiral limit is simultaneously the explanation for a proton mass of around 1\,GeV.  Thus, enigmatically, properties of the nearly-massless pion are the cleanest expression of the mechanism that is responsible for almost all the visible mass in the Universe.  There are numerous, additional corollaries, some of which are described in Refs.\,\cite{Holl:2005vu, Chang:2012cc, Li:2016dzvS}; but, most importantly, no approach that is incapable of expressing Eq.\,\eqref{gtrE} can truly claim to provide a veracious understanding of pion structure.

At this point, it is worth returning to Eq.\,\eqref{anomalypion}.  The pion's Poincar\'e-invariant mass and Poincar\'e-covariant wave function are obtained by solving a Bethe-Salpeter equation.  This is a scattering problem.  In the chiral limit, two massless fermions interact via exchange of massless gluons, \emph{i.e}.\ the initial system is massless; and it remains massless at every order in perturbation theory.  The complete calculation of the scattering process, however, involves an enumerable infinity of dressings and scatterings.  This can be represented by a coupled set of gap- and Bethe-Salpeter equations.  At a renormalisation scale typical of hadronic phenomena, $\zeta = 2\,{\rm GeV}=:\zeta_2$, it is practical to build the kernels using a dressed-parton basis, \emph{viz}.\ from valence-quarks with a momentum-dependent running mass produced by self-interacting gluons, which have given themselves a running mass.

In the chiral limit one can prove algebraically \cite{Munczek:1994zz, Bender:1996bb, Chang:2009zb, Binosi:2016rxz} that, at any finite order in a symmetry-preserving construction of the kernels for the gap (quark dressing) and Bethe-Salpeter (bound-state) equations, there is a precise cancellation between the mass-generating effect of dressing the valence-quarks and the attraction introduced by the scattering events.  This cancellation guarantees that the simple system,  which began massless, becomes a complex system, with a nontrivial bound-state wave function that is attached to a pole in the scattering matrix, which remains at $P^2=0$, \emph{i.e}. the bound-state is also, therefore, massless.

The precise statement is that in the pseudoscalar channel, the dynamically generated mass of the two fermions is precisely cancelled by the attractive interactions between them, if and only if, Eq.\,\eqref{gtrE} and its companions \cite{Maris:1997hd, Qin:2014vya, Binosi:2016rxz} are satisfied.
It can be expressed as follows \cite{Roberts:2016vyn}:
\begin{subequations}
\label{QuasiParticle}
\begin{align}
\langle \pi(q) | \theta_0 | \pi(q) \rangle
& \stackrel{\zeta \gg \zeta_2}{=} \langle \pi(q) | \tfrac{1}{4} \beta(\alpha(\zeta)) G^{a}_{\mu\nu}G^{a}_{\mu\nu} | \pi(q) \rangle  \stackrel{\zeta \simeq \zeta_2}{\to}
\langle \pi(q) | {\cal D}_1 + {\cal I}_2 |\pi(q)\rangle\,, \\
{\cal D}_1 & = \sum_{f=u,d} M_f(\zeta) \, \bar {\cal Q}_f(\zeta) {\cal Q}_f(\zeta) \,, \quad
{\cal I}_2   = \tfrac{1}{4} [\beta(\alpha(\zeta)) {\cal G}^{a}_{\mu\nu}{\cal G}^{a}_{\mu\nu}]_{2{\rm PI}}  \,,
\end{align}
\end{subequations}
which describes the metamorphosis of the parton-basis chiral-limit expression of the expectation-value of the trace anomaly into a new expression, written in terms of a nonperturbatively-dressed quasi-particle basis, with dressed-quarks denoted by ${\cal Q}$ and the dressed-gluon field strength tensor by ${\cal G}$.
Here, the first term is positive, expressing the one-body-dressing content of the trace anomaly.  Plainly, a massless valence-quark (antiquark) acquiring a large mass through interactions with its own gluon field is an expression of the trace-anomaly in what might be termed the one-quasiparticle subsector of a complete pion wave function.
The second term is negative, expressing the two-particle-irreducible (2PI) scattering-event content of the forward scattering process represented by this expectation-value of the scale-anomaly.  This term acquires a scale because the couplings, and the gluon- and quark-propagators in the 2PI processes have all acquired a mass-scale.
Away from the chiral limit, and in other channels, such as the proton, the cancellation is incomplete. 

It is worth reiterating that these statements can be verified algebraically.  Exact cancellation for the pion in the chiral limit was first demonstrated in Ref.\,\cite{Nambu:1961tp}; and in connection with the gap and pseudoscalar-channel Bethe-Salpeter equations, Ref.\,\cite{Roberts:2010rn} provides the simplest realisation.  The same framework yields algebraic formulae for the masses of the nucleon and $\Delta$-baryon, showing that these states possess masses on the order of the three-times the DCSB mass-scale ($\sim 3 M(0)$) in the chiral limit \cite{Roberts:2011cf} despite the pion's masslessness.  The most sophisticated continuum analyses of the pion and proton masses may be traced from Refs.\,\cite{Eichmann:2008ae, Eichmann:2008ef, Chang:2009zb}.

The fact that a Poincar\'e-invariant analysis of the simultaneous impact of the trace anomaly in the pion and proton cannot yield a sum of terms that is each individually positive precludes a useful ``pie diagram'' breakdown of the distribution of mass within a hadron.

\begin{figure}[t]
\begin{minipage}[t]{0.95\textwidth}
\begin{minipage}{0.5\textwidth}
\centerline{\includegraphics[clip,width=0.97\textwidth]{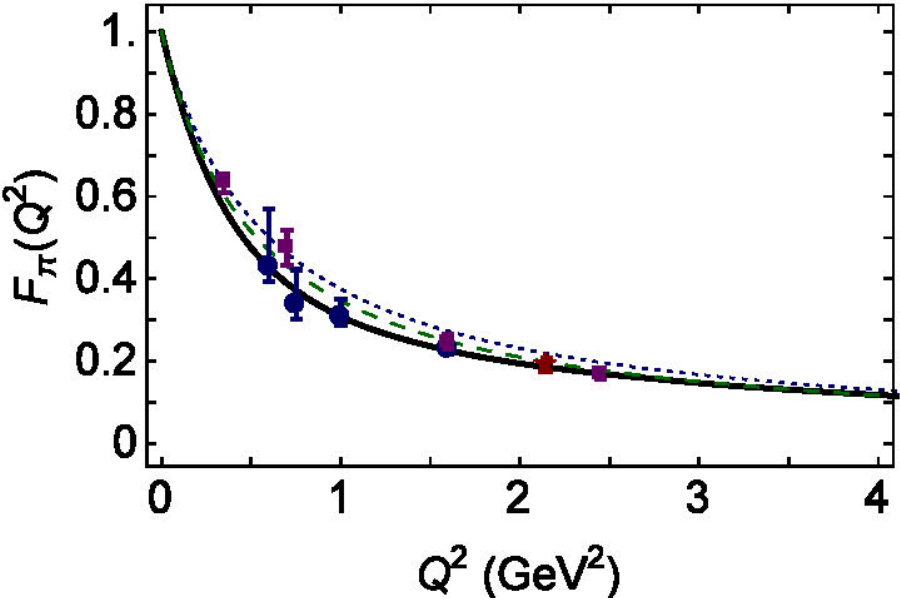}}
\end{minipage}
\begin{minipage}{0.5\textwidth}
\centerline{\hspace*{2em}\includegraphics[clip,width=0.97\textwidth]{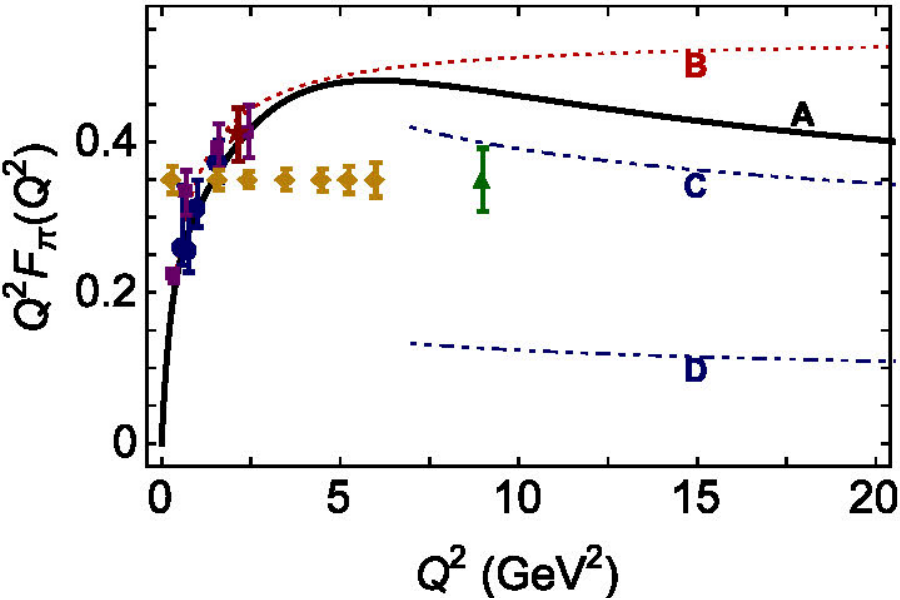}}
\end{minipage}
\end{minipage}
\caption{\label{Fig2}
\textbf{Left panel}.
\emph{Solid curve} -- Charged pion form factor computed in Ref.\,\cite{Chang:2013niaS} ($r_\pi = 0.66\,$fm \emph{cf}.\ experiment \cite{Agashe:2014kda} $r_\pi = 0.672 \pm 0.008\,$fm); \emph{long-dashed curve} -- calculation in Ref.\,\protect\cite{Maris:2000sk}; and \emph{dotted curve} -- monopole form ``$1/(1+Q^2/m_\rho^2)$,'' where $m_\rho=0.775\,$GeV is the $\rho$-meson mass.  In both panels, the filled-star is the point from Ref.\,\cite{Horn:2007ug}, and the filled-circles and -squares are the data described in Ref.\,\protect\cite{Huber:2008id}.
\textbf{Right panel}.
$Q^2 F_\pi(Q^2)$.  \emph{Solid curve}\,(A) -- prediction in Ref.\,\cite{Chang:2013niaS}.
Remaining curves, from top to bottom: \emph{dotted curve}\,(B) -- monopole form fitted to data in Ref.\,\protect\cite{Amendolia:1986wj}, with mass-scale $0.74\,$GeV;
\emph{dot-dot--dashed curve}\,(C) -- perturbative QCD (pQCD) prediction obtained using the modern, dilated pion PDA from Ref.\,\cite{Chang:2013pqS};
and \emph{dot-dot--dashed curve}\,(D) --  pQCD prediction computed with the conformal-limit PDA, $\varphi_\pi^{\rm cl}(x)$, which had previously been used to guide expectations for the
asymptotic behaviour of $Q^2 F_\pi(Q^2)$.
The filled diamonds and triangle indicate the projected reach and accuracy of forthcoming experiments  \protect\cite{E1206101, E1207105}.}
\end{figure}

\section{Hadron observables}
As stated at the outset, physics is an empirical science.  Consequently, one should ask for observable predictions following from the ideas presented above, with empirical verification if they are to be taken seriously.  It is important, therefore, that methods have recently been developed which enable one to project the pion's Poincar\'e-covariant Bethe-Salpeter wave function, onto the light front \cite{Chang:2013pqS}.  This provides, \emph{inter alia}, the pion's leading-twist parton distribution amplitude (PDA), which is the closest thing in QCD to a quantum mechanical wave function for the pion.  At an hadronic scale, the computed result is concave and significantly broader than the asymptotic distribution amplitude, $\varphi_\pi^{\rm cl}(x)=6 x(1-x)$ \cite{Lepage:1979zb, Efremov:1979qk, Lepage:1980fj}.  This establishes that $\varphi_\pi(x)$ is hardened at an hadronic scale as a direct consequence of DCSB.

How does this compare with results from simulations of lQCD?  Well, with current algorithms, lQCD can only determine one non-trivial moment of the pion PDA.  However, the methods developed in Refs.\,\cite{Chang:2013pqS, Cloet:2013ttaS} are powerful enough to extract physics from that limited input.  In fact, they enable the full $x$-dependence of meson PDAs to be inferred from the limited lQCD output, revealing complete consistency between the continuum predictions and the lattice results \cite{Shi:2014uwaS, Shi:2015esaS, Horn:2016rip}.

Armed with a reliable prediction of the pion's PDA, one can validate the collinear hard-scattering formula for the pion's elastic electromagnetic form factor, $F_\pi(Q^2)$, \emph{if} a scheme is available that enables direct computation of $F_\pi(Q^2)$ on the entire domain of spacelike momenta.  Such a method was introduced in Ref.\,\cite{Chang:2013niaS}; and this enabled a demonstration of the fact that the leading-order, leading-twist pQCD result for $Q^2 F_\pi(Q^2)$ underestimates the full computation by just 15\% on $Q^2 \gtrsim 8\,$GeV$^2$ (see Fig.\,\ref{Fig2}), in stark contrast to the result obtained with $\varphi_\pi^{\text{cl}}(x)$.  The analysis shows that hard contributions to $F_\pi(Q^2)$ dominate for $Q^2 \gtrsim 8\,$GeV$^2$; but, even so, the magnitude of $Q^2 F_\pi(Q^2)$ reflects the scale of DCSB, a pivotal emergent phenomenon in the standard model.  It follows that a measurement of the pion form factor on $Q^2 \gtrsim 8\,$GeV$^2$ should expose parton model scaling and QCD scaling violations and thereby achieve a goal that was a driving force for the construction of CEBAF.

\begin{figure}[t]
\centering
\sidecaption
\includegraphics[width=7cm,clip]{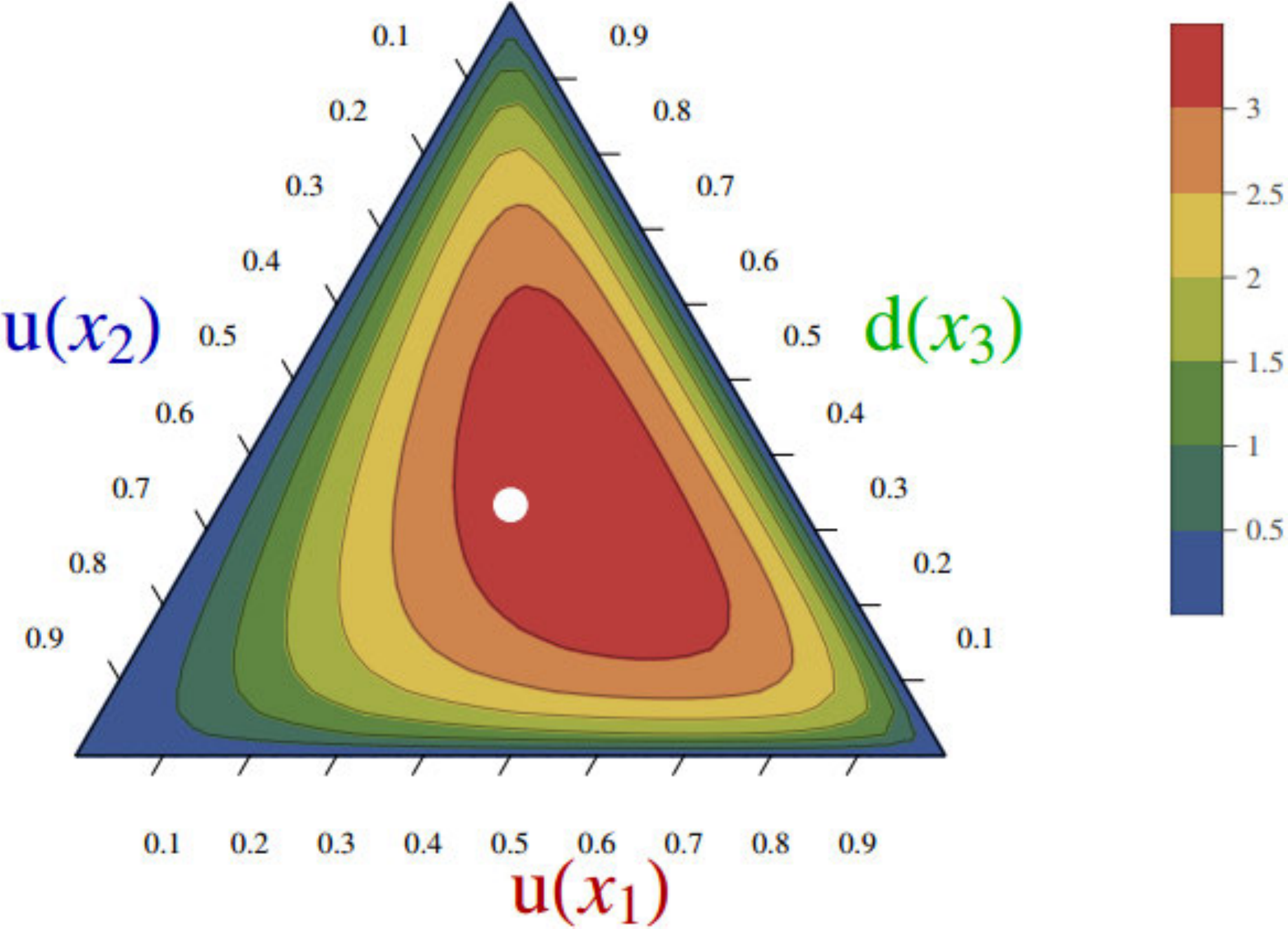}
\caption{\label{FigCMezrag}
Barycentric plot of a leading-twist parton distribution amplitude (PDA), $\varphi_p(x_1,x_2,x_3)$, for a proton containing both scalar and axial-vector diquark correlations, with the scalar diquark alone contributing 62\% to the bound-state's normalisation, as is found in realistic computations \cite{Segovia:2014aza}.  In the absence of correlations, this PDA would be a symmetric function, peaked at $x_1=x_2=x_3=\tfrac{1}{3}$, the point marked by the open circle.  The PDA's distortion matches that inferred from simulations of lQCD \cite{Anikin:2013aka}, and owes directly to the presence of non-pointlike diquark scalar and axial-vector diquark correlations.
}
\end{figure}

There are many more verifiable predictions of meson properties, \emph{e.g}.\ the light meson spectrum \cite{Chang:2011ei}; the parton distribution functions of light-quark mesons \cite{Chang:2014lvaS, Mezrag:2014jkaS, Chen:2016sno, Mezrag:2016hnp} and distribution amplitudes of heavy-heavy mesons \cite{Ding:2015rkn, Li:2016mah}; and numerous elastic and transition form factors \cite{Raya:2015gvaS, Raya:2016yuj}.  However, baryons are, at least, equally interesting; and there is a striking prediction for baryon structure driven by DCSB, \emph{viz}.\ they contain strong, non-pointlike diquark correlations, as illustrated in Fig.\,\ref{FigCMezrag}, the existence of which has numerous observable consequences, as explicated, \emph{e.g}.\ in Refs.\,\cite{Cloet:2013gva, Roberts:2013mja, Segovia:2014aza, Segovia:2015hraS, Segovia:2015ufa, Roberts:2016dnb, Segovia:2016zyc}.

\section{Epilogue}
In drawing this contribution to a close, it is worth reiterating a few points.

Owing to the conformal anomaly, both gluons and quarks acquire mass dynamically in QCD.  Those masses are momentum dependent, with large values at infrared momenta: $m(k^2\simeq 0) > \Lambda_{\rm QCD}$.
The appearance of these nonperturbative running masses is intimately connected with confinement and DCSB; and the relationship between those phenomena entails that in a Universe with light-quarks, confinement is a dynamical phenomenon.  Consequently, static-quark flux tubes are not the correct paradigm for confinement and it is practically meaningless to speak of linear potentials and Regge trajectories in connection with observable properties of light-quark hadrons.
Moreover, the origin and distribution of a hadron's mass depends on the observer's preferred frame of reference and resolving scale, and for contemporary and planned experiments, the DCSB paradigm is the best way to explicate and understand the associated, emergent phenomena.

In exploring the connection between QCD's gauge and matter sectors, top-down and bottom-up analyses in continuum-QCD have converged on the form of the renormalisation-group-invariant interaction.  This outcome paves the way to parameter-free predictions of hadron properties; and, combined with decades of studying the three valence-body problem in QCD, provides the evidence necessary to conclude that diquark correlations are a reality.  Diquarks are complex objects, however, so their existence does not restrict the number of baryon states in any obvious way.  This effort has led to a sophisticated understanding of the nucleon, $\Delta$-baryon and Roper resonance: all may be viewed as Borromean bound-states, and the Roper is at heart the nucleon's first radial excitation.

The material presented herein highlights the capacity of continuum methods in QCD to connect the quark-quark interaction, expressed, for instance, in the dressed-quark mass function, $M(p^2)$, with predictions for a wide range of hadron observables.  It therefore serves as strong motivation for new experimental studies of, \emph{inter alia},
elastic and transition form factors, and parton distribution amplitudes and functions, for both gluons and quarks,
which exploit the full capacity of modern and planned facilities in order to chart the running coupling and masses of QCD and thereby explain the origin of more than 98\% of the visible mass in the Universe.
This must shed light on confinement, one of the most fundamental problems in modern physics and a puzzle whose solution is unlikely to be found in a timely fashion through theoretical analyses alone.  A multipronged approach is required, capitalising on the empirical nature of physics, and hence involving constructive feedback between experiment and theory.

\section*{Acknowledgments}
Both the results described and the insights drawn herein are fruits from collaborations we have joined with many colleagues and friends throughout the world; and we are very grateful to them all.
CDR would also like to thank the organisers and their committees for enabling his participation in the XIIth Quark Confinement and the Hadron Spectrum, 29 August - 3 September 2016, Thessaloniki, Greece, which proved very rewarding.
This work was supported by the U.S.\ Department of Energy, Office of Science, Office of Nuclear Physics, under contract no.~DE-AC02-06CH11357.

%
%
%


%
%


\end{document}